\RequirePackage{rotating}
\documentclass[a4paper,fleqn,usenatbib]{mnras}

\usepackage[T1]{fontenc}
\usepackage{ae,aecompl}

\usepackage{rotating}
\usepackage{graphicx}	
\usepackage{amsmath}	
\usepackage{amssymb}	
\usepackage{caption}
\usepackage{pdflscape}

\title[Radio emission from magnetic OB stars]{A JVLA survey of the high frequency radio emission of the massive magnetic B- and O-type stars}

\author[Kurapati et al.]{
Sushma Kurapati,$^{1}$\thanks{E-mail: sushma@ncra.tifr.res.in}
Poonam Chandra,$^{1}$, 
Gregg Wade $^{2}$, 
David H. Cohen$^{3}$, 
\newauthor
Alexandre David-Uraz$^{4}$, 
Marc Gagne$^{5}$, 
Jason Grunhut$^{6}$, 
Mary E. Oksala$^{7,8}$, 
\newauthor
Veronique Petit$^{4}$, 
Matt Shultz$^{2,9}$, 
Jon Sundqvist$^{10,11}$, 
Richard H. D. Townsend$^{12}$, 
\newauthor
Asif ud-Doula$^{13}$\\
$^{1}$ National Centre for Radio Astrophysics, Tata Institute of Fundamental Research, PO Box 3, Pune 411007, India\\
$^{2}$ Department of Physics, Royal Military College of Canada, PO Box 17000, Station Forces, Kingston, Ontario K7K 7B4, Canada\\
$^{3}$ Department of Physics and Astronomy, Swarthmore College, Swarthmore, PA 19081, USA\\
$^{4}$  Department of Physics \& Space Sciences, Florida Institute of Technology, Melbourne, FL 32901, USA\\
$^{5}$ Dept. of Earth \& Space Sciences, West Chester University, West Chester, PA 19383 \\
$^{6}$ Dunlap Institute for Astronomy and Astrophysics, University of Toronto, Rm 101, 50 St. George Street, Toronto, ON, M5S 3H4, Canada\\
$^{7}$ Department of Physics, California Lutheran University, 60 West Olsen Road \#3700, Thousand Oaks, CA, 91360, USA\\
$^{8}$ LESIA, Observatoire de Paris, PSL Research University, CNRS, Sorbonne Universit\'es, UPMC Univ. Paris 06, F-92195 Meudon, France\\
$^{9}$ Department of Physics, Engineering Physics and Astronomy, Queen's University, 99 University Avenue, Kingston, ON K7L 3N6, Canada\\
$^{10}$ KU Leuven, Instituut voor Sterrenkunde, Celestijnenlaan 200D, 3001 Leuven, Belgium\\
$^{11}$ Centro de Astrobiologia, CSIC-INTA, Departamento de Astrofisica, Ctra. Torrejon a Ajalvir km.428850 Madrid, Spain \\
$^{12}$ Department of Astronomy, 2535 Sterling Hall 475 N. Charter Street, Madison, WI 53706-1582, USA\\
$^{13}$ Penn State Worthington Scranton, 120 Ridge View Drive, Dunmore, PA 18512, USA
}

\date{Accepted XXX. Received YYY; in original form ZZZ}

\pubyear{2016}

\begin{document}
\label{firstpage}
\pagerange{\pageref{firstpage}--\pageref{lastpage}}
\maketitle

\begin{abstract}
We conducted a survey of seven magnetic O and eleven B-type stars with masses above
$8M_\odot$
using the Very Large Array in the 1cm, 3cm and 13cm bands.
The survey resulted in a detection of two O and two B-type stars.
While the detected O-type stars - HD\,37742 and HD\,47129 - are in binary systems, the detected B-type stars, HD\,156424 and ALS\,9522, are not known to be in binaries.
All four stars were detected at 3cm, whereas three were detected at 1cm and only one star was detected at 13cm.
The detected B-type stars are significantly more radio luminous than the non-detected ones, which
is not the case for O-type stars.
The non-detections at 13cm are interpreted as due to thermal free-free absorption.
Mass-loss rates were estimated using 3cm flux densities and were compared with theoretical mass-loss rates,
which assume free-free emission. For HD\,37742,
the two values of the mass-loss rates were in good agreement,
possibly suggesting that the radio emission for this star is mainly thermal. For the other three stars, the estimated mass-loss rates from radio observations were 
much higher than those expected from theory, suggesting either a possible contribution from 
non-thermal emission from the magnetic star or thermal or non-thermal emission due to interacting winds of the binary system, 
especially for HD\,47129. All the detected stars are predicted to host centrifugal magnetospheres except HD\,37742, which is likely to host a dynamical magnetosphere. This suggests that non-thermal radio emission is favoured in stars with centrifugal magnetospheres.

\end{abstract}




\section{Introduction}


Recent systematic surveys of the magnetic properties of hot stars \citep[MiMeS and BoB surveys; ][]{wade16, morel14} have revealed a  population of O- and B-type stars hosting significant surface magnetic fields \citep{petit13,shultz}.  These magnetic fields are strong (from a few hundred to tens of thousands of gauss), organised (mainly dipolar) and highly stable. 

Theoretical models and magnetohydrodynamical (MHD) simulations have explored the dynamical 
interaction of these magnetic fields with stellar rotation and mass-loss \citep[e.g. ][]{uddoula02, owocki08}. Both observation and theory show clearly that the stellar wind interaction with the magnetic field leads to wind confinement and channelling, generating a long-lived circumstellar magnetosphere \citep[e.g.][]{shore90, babel97}. Close to the star, the magnetic pressure dominates the kinetic 
pressure of the wind, forcing the wind to follow closed magnetic field lines in regions near the magnetic equatorial plane. Far from the star, the kinetic pressure dominates the magnetic pressure due 
to the stronger decline of the magnetic energy density compared to the wind kinetic energy density. The radius at which the energy densities
become equal is defined as the Alfv\'en radius, which also marks the boundary of  the inner magnetosphere.  
Beyond the Alfv\'en radius, the stellar wind
opens the magnetic field lines and generates a current sheet in the magnetic equatorial plane. This region is  the middle magnetosphere,
where the electrons are accelerated to relativistic speeds and gyrosynchrotron radio emission is expected to arise  \citep{trigilio04}.  The gyrosynchrotron
emitting region extent is defined by these wind electrons returning to the  star along the field lines.

\citet{petit13} classified the magnetospheres into two broad physical categories, namely dynamical magnetospheres (DM) and centrifugal magnetospheres (CM). 
This classification is based on two parameters: the degree of magnetic wind confinement characterized by the Alfv\'en radius ($R_{\rm A}$), and stellar rotation 
characterized by the Kepler co-rotation radius ($ R_{\rm K}$).  A DM results in the case of a slowly rotating star ($ R_{\rm A} < R_{\rm K}$). In this case, wind plasma 
trapped in closed magnetic loops falls back onto the stellar surface on a dynamical timescale. A CM occurs in the case of a rapidly 
rotating star ( $ R_{\rm A} > R_{\rm K}$). In this case, wind plasma caught in the region between $ R_{\rm A} $ and $ R_{\rm K} $ is centrifugally supported against infall. This results in long-term accumulation of plasma and consequently higher magnetospheric plasma density.

In stars with high mass-loss rates, radio emission is usually produced by thermal free-free emission from the ionised stellar wind. However, as discussed above, non-thermal synchrotron emission may dominate the radio spectrum in the presence of magnetic field and relativistic electrons. Electrons can be accelerated to relativistic speeds, either by magnetic reconnection near the current sheet in the middle magnetosphere \citep{um92} or through Fermi acceleration in strong shocks in the inner magnetosphere \citep{owocki84, eich93}. Thermal radio emission, X-ray emission and optical line emission are produced from  the inner magnetosphere, whereas non-thermal radio emission comes from the middle magnetosphere. Hence, non-thermal radio emission probes different regions of the magnetosphere. In addition, as the emitting plasma is optically thick at long wavelengths, radio radiation at different frequencies probes the stellar magnetosphere at different depths, and from different perspectives as the star rotates. { As a consequence, magnetic stars show radio flux variability with the stellar rotation period. The rotation modulation in radio emission was first suggested by \citet{Leone91}  for
a rigidly rotating magnetosphere. These modulations have been observed in the radio light curves of some magnetic B-A stars  \citep[e.g.][]{bailey12,chandra15}
and 
have been then modeled by \citet{trigilio04, Leto06, leto12}. }
 
Some numerical simulations \citep{vanloo05} suggest that one requires both a magnetic field and a binary companion  to explain the non-thermal radio 
emission from massive stars. In these simulations, the synchrotron radiation from single massive stars is produced relatively close to the star. Due to the 
large free-free opacity of the stellar winds at radio wavelengths, radiation emitted too close to the star will be absorbed { \citep{Blomme11}}. When a star has a massive nearby 
binary companion (orbital period of a few days), emission is produced from the wind collision region where shocks occur,{{ which can escape due to a 
lower free-free opacity \citep{vanloo05}}}. Thus the observed radio flux is likely to have a contribution due to the colliding wind  in addition to the 
star's magnetosphere and thermal free-free emission from the star.
 
Various investigations have been carried out to observe the radio emission of magnetic A- and B-type stars \citep[e.g. ][]{abbott85, drake87, linsky92}. 
However, no such studies have been made of  the radio properties of hotter magnetic B- and O-type stars. 
\citet{bieging89} carried out a Very Large Array (VLA) survey on 88 O-type and early B-type stars, which are not known to be magnetic. They detected a total 
of 14 sources at 6cm with six of them displaying non-thermal
radio emission. They concluded that non-thermal radio emission is more common in very luminous stars. 


To homogenize the study of the physics of magnetospheres of hot stars, we are carrying out a systematic survey of the radio emission properties of the known magnetic O and B type stars at high frequencies using the Karl J. Jansky Very
 Large Array (JVLA)  and at low frequencies using the Giant Metrewave Radio Telescope (GMRT). This paper reports the high frequency JVLA observations. \citet{chandra15} reported GMRT observations of 8 magnetic B type stars and 3 magnetic O type stars. Out of the eight B-type stars, five were detected in both the 1390 and 610 MHz bands. The three O-type stars were observed only in the 1390 MHz band, and no detections were obtained. This result was explained as a consequence of free-free absorption by the freely flowing stellar wind exterior to the confined magnetosphere. The low frequency GMRT survey thus puts strong constraints on 
 free-free absorption of radio emission, and combined with high frequency VLA data can provide a more complete description of  the characteristics of the 
 medium  surrounding the star.

 In this paper, we present results of JVLA observations of 18 known magnetic O- and B- type stars with masses $M \ge 8 M_\odot$. In \S \ref{obs}, we describe the observations and data analysis.  In \S \ref{results}, we report the detections and the mass-loss rates that are calculated from radio fluxes. In \S \ref{discussion}, we compare the properties of detected and non-detected stars and explore their properties. In \S \ref{summary}, we present our conclusions. 

\section{Observations $ \& $  Data analysis}
\label{obs}

%
  %

\subsection{Sample}

The basis for target selection is the complete list of known magnetic OB stars compiled by \citet{petit13}. From this list, we have selected all the stars with mass $\geq 8 M_{\odot} $ and declination above $-40$ degrees. This sample consists of seven O- type stars and eleven B-type stars. The list of targets, along with their stellar parameters from \citet{petit13}, \citet{naze14}, and \citet{shultz}, are given in Table \ref{table1}.

\begin{sidewaystable*}
     \centering
%
\caption{Stellar and magnetic properties (spectral type, distance, luminosity, effective temperature, mass, radius, polar strength of the magnetic field, rotation period, Alfv\'en radius, Kepler corotation radius and wind terminal velocity) for the sample of magnetic O- and B-type stars. }
\label{table1}
\begin{tabular}{ p{0.5cm} p{1.65cm} p{1cm} p{1.2cm} p{0.55cm} p{1.0cm}  p{0.85cm} p{0.55cm} p{0.65cm} p{0.55cm} p{0.7cm} p{0.65cm} p{0.65cm} p{0.9cm} p{1.0cm} }
\\
\hline
 ID & Name & Binary Comp. & Sp.type & $d$ (pc) & log($L _{\ast}$) (L$ _{\odot} $) & $T_{\rm eff}$ (kK) & $ M_{\ast} $ (M$_{\odot} $) & $ R_{\ast} $ (R$_{\odot} $) & $ B_{\rm p} $ (kG)  & Period (days)  & $ R_{\rm A} $  (R$_{\ast} $) & $ R_{\rm K} $  (R$_{\ast} $) & $ v_{\infty} $ (kms$ ^{-1} $) & Binary status \\ 
\hline
 1 & HD\,191612 & Primary &Of$ ? $p & 2290 & 5.4$ \pm $0.2 & 35$ \pm $1 & 30 & 14 &  $  $2.5 & 537.2 & 3.7 & 57 & 2119 & BIN \\
     & & Secondary & B1 & - & - & 20 & 15 & -  & -& -& -& -&- & \\ 
     \hline
 2 & NGC\,1624-2 & - & Of$ ? $cp & 5152 & 5.1$ \pm $0.2 & 35$ \pm $2 & 34 & 9.7 & $ > $20 & 158 & $ > $11 & 41 & 2890 & - \\
   \hline
 3 & CPD\,-282561 & - & Of$ ? $p & 6100 & 5.5$ \pm $0.2 & 35$ \pm $2 & 43 & 14 & $ > $1.7 & 73.4 & 3.4 & 18.7 &2400 & - \\ 
    \hline
 4 & HD\,47129 & Secondary &O7.5 III & 1584 & 5.31$ \pm $0.03 & 37$ \pm $2 & 22 & 11 & $ > $2.8 & 1.21551 & $ > $5.4 & $ < $2.2 & 3567 & BIN \\ 
      & & Primary & O8 III/I & -& 5.14$\pm $0.04 & 32$\pm $2  & 3 & 12  & -& -& -& -&- & \\
         \hline
 5 & HD\,108 & - & Of$ ? $p & 2510 & 5.7$ \pm $0.1 & 35$ \pm $2 & 43 & 19 & $ > $0.50 & 18000 & $ > $1.7 &526 & 2022 & - \\ 
         \hline
 6 & HD\,57682 &- & O9 V & 1300 & 4.8$ \pm $0.2 & 34$ \pm $1 & 17 & 7 & 1.7 & 63.571 & 3.7 &24 &2395 & - \\
    \hline
     7 & HD\,37742 & Primary & O9.5 Ib & 414 & 5.6$ \pm $0.1 & 29.5$ \pm $1 & 33 & 20 & 0.140 & 7 & 1.1 &2.1 &1723 & BIN \\
      & &Secondary & B1 IV & - & 4.4 & 29 & 14 & 7  & -& -& -& -&- & \\
         \hline
 8 & HD\,149438 &- & B0.2 V & 180 & 4.5$ \pm $0.1 & 32$ \pm $1 & 17.2 & 6.0 & 0.2 & 41.033 & 2.0 & 20.3 & 2581 & - \\
    \hline
 9 & HD\,66665 &- & B0.5 V & 1500 & 4.2$ \pm $0.5 & 28$ \pm $1 & 9 & 5.5 & 0.67 & 21 & 3.1 & 12.3 & 2256 & - \\
    \hline
 10 & HD\,46328 & - &B1 III & 423 & 4.6$ \pm $0.1 & 27$ \pm $2 & 13.4 & 7.5 & $ > $1.5 & 8600 & 5.7 & 867.4 & 1984 & - \\
    \hline
 11 & ALS\,8988 & - &B1 & 1880 & 4.1$ \pm $0.1 & 27$ \pm $1 & 12 & 4.7 & $ > $1.5 & - &$ > $7.5  &$ < $9.6 & & -\\
    \hline
 12 & HD\,47777 &- & B1 III & 760 & 4.1$ \pm $0.1 & 27$ \pm $2 & 9 & 5 & $ > $2.1 & 537.2 & $ > $8.6 &$ < $4.3 &2142 & - \\
    \hline
 13 & HD\,205021 & Primary & B1 IV & 182 & 4.2$ \pm $0.1 & 26$ \pm $1 & 12 & 6.5 & 0.36 & 12.001 & 3.1 & 7.3 & 2064 & BIN \\
      &- &Secondary & B5 Ve & - & - & 17 & 4 & -  & -& -& -& -&- & \\
         \hline
 14 & HD\,163472 &- & B1/2 V & 290 & 3.8$ \pm $0.1 & 25$ \pm $1 & 10.2 & 4.6 & 0.40 & 3.639 & 7.3 & 4.5 & 2344 & - \\
    \hline
 15 & ALS\,9522 &- & B1.5 V & 1800 & 4.0$ \pm $0.1 & 22$ \pm $2 & 10 & 6.4 & $ > $4.0 & - & $ > $11 &$ < $2 & 989 & - \\
    \hline
 16 & HD\,156424 &- & B2 V & 1100 & 3.7$ \pm $0.4 & 22$ \pm $3 & 7.4 & 5.0 & $ > $0.65 & 2.87 & 15.7 & 3.1 & 957 & - \\
    \hline
 17 & HD\,3360 & - &B2 IV & 183 & 3.7$ \pm $0.2 & 20.4$ \pm $0.9 & 8.2 & 5.7 & $ > $0.34 & 5.371 & 2.5 & 4.2 & 932 & -\\
    \hline
 18 & HD\,58260 &- & B3 Vp & 826 & 4.1$ \pm $0.2 & 20$ \pm $2 & 8.8 & 9.5 & $ > $7.0 & - & 14.4 & 9.9 & 766 & -\\
 \hline
 \multicolumn{15}{l}{Notes: Taken from  \citet{petit13}, \citet{naze14}, \citet{hum13}, \citet{wheel09} and \citet{shultz} $\&$ ref therein.} 

\end{tabular}


\end{sidewaystable*}

\subsection{Observations}

The JVLA observations were taken between 2014 March 8 to 2014 August 1 during the 14A semester in the 13\,cm (S),
3\,cm (X) and 1\,cm (Ka) bands. 
The data were collected in an 8-bit sampler mode for the S band, and in a 3-bit sampler mode for the X and Ka bands. Thus the bandwidths for
S, X and Ka bands observations were 2 GHz (frequency range 2--4\,GHz ), 4\,GHz (frequency range 8--12\,GHz ) and 
8\,GHz (frequency range 29--37\,GHz), respectively.  The observations were taken in VLA A, D and A$\rightarrow$D configurations. 
The duration of each observation, including overheads, in S and X bands was 30\,minutes
and 1\,hour for the Ka band.  For each observation, a flux calibrator and a phase calibrator were observed along with the target source. The flux calibrator was observed once, either at the start or at the end of the observation. The phase calibrator was observed once every five minutes for the S and X bands and once every three minutes for the Ka band observations. Approximately fifteen minutes was spent on the target source for S band and X band observations and thirty minutes was spent on the target source for Ka band observations.

\subsection{Data analysis}

All the calibration and data reduction were carried out using standard tasks in the Common Astronomy Software Applications (CASA) Package. The initial reduction steps such as flagging bad data, correcting for atmospheric opacities, antenna delay solutions and bandpass corrections were done using the VLA calibration pipeline. In some cases, additional flagging and recalibration was required after the pipeline calibration. In such cases, flagging was done manually and the pipeline was used again to recalibrate the data. The flagged and calibrated visibility data were used to produce continuum images using the {\sc clean} algorithm in CASA. Wide-band effects were taken into account using multi-frequency synthesis, where two Taylor coefficients were used to model the sky frequency dependence. All the detected sources were unresolved at JVLA frequencies. Flux densities on these radio sources were measured by fitting elliptical gaussian functions to them using the CASA task {\sc IMFIT}. Both the uncertainities in the Gaussian fit and the local rms noise were taken into account for calculating the uncertainity in measured flux density. The rms noise, flux densities for the detected sources and 3$ \sigma$ upper limits for the non-detections are included in Table \ref{table2}.

\begin{sidewaystable*}
\centering
\caption{VLA Observations of magnetic O-type and B-type stars }
\label{table2}
\begin{tabular}{ p{1.65cm} p{1.38cm} p{1.45cm} p{0.5cm} p{0.6cm}  p{1.38cm} p{1.45cm} p{0.5cm} p{0.6cm} p{1.38cm} p{1.45cm} p{0.5cm} p{0.6cm} }
\\
\hline
Name &  \multicolumn{4}{c}{13cm} & \multicolumn{4}{c}{3cm} & \multicolumn{4}{c}{1cm} \\
\hline
 & Obs.Date  & Flux density (mJy) & rms ($ \mu $Jy)& Mean Phase & Obs.Date  & Flux  density (mJy) & rms ($ \mu $Jy)& Mean Phase & Obs.Date  & Flux  density (mJy) & rms ($ \mu $Jy)& Mean Phase \\
\hline
HD\,191612 & Apr 25.74& $ < $0.037& 12& 0.17 & Apr 23.47 &$ < $ 0.019&6.5& 0.24 & May 02.60 & $ < $0.031& 10.3 & 0.26 \\ 
 NGC\,1624-2 & May 05.95 &$ < $0.024 & 8& 0.17 & Apr 29.90 &$ < $ 0.018 & 6.2& 0.13 & Aug 01.63& $ < $0.042& 14 & 0.72 \\
 CPD\,-282561 & May 05.97 &$ < $0.028 & 9.5 & 0.12 & Apr 29.11 &$ < $ 0.021 & 7& 0.12& - & - & - & - \\ 
 HD\,47129& May 10.91& $ < $0.03& 10& 0.71& May 06.97 & 0.208$ \pm $0.015  &7.2& 0.47& Jul 13.64 & 0.392$ \pm $0.037& 13.0 & 0.13 \\ 
 HD\,108 &- &- &- & - & May 03.92& $ < $0.019 & 6.3& - & Jul 12.62& $ < $0.038& 12.6 & -  \\ 
 HD\,57682 & Apr 04.87&$ < $0.024 & 8.2& 0.54 & Apr 03.95 &$ < $ 0.029 & 9.8& 0.54 & Apr 05.21& $ < $0.030& 10 & 0.56 \\
 HD\,37742 & Apr 17.01& 0.494$ \pm $0.025 &8.3& 0.64& Apr 16.03& 1.104$ \pm $0.015 &6 & 0.50& Jul 29.57 & 2.765$ \pm $0.049& 15.0& 0.43 \\
 HD\,149438 & Apr 29.3 & $ < $36 &12& - & Apr 25.29&$ < $ 0.020 & 6.5& - & May 11.28& $ < $0.035& 11.7 & - \\
 HD\,66665 & Apr 30.17&  $ < $42 &14& 0.35 & Apr 29.20 & $ < $ 0.018 & 6 & 0.31 & May 09.16& $ < $0.033& 11.0& 0.72  \\
 HD\,46328 & Apr 03.99& $ < $ 45& 15& 0.14 & Apr 03.97 &$ < $ 0.018 & 6& 0.14 & Jul 19.64 & $ < $0.048& 16.0 & 0.16  \\
 ALS\,8988 & May 12.98& $ < $30 & 10& - & Apr 29.92 & $ < $ 0.020 &6.7& - & Jul 11.67 & $ < $0.044& 14.8 & - \\
 HD\,47777 & - & - & - & - & Apr 07.90 &$ < $ 0.036 & 12& - & Apr 08.22 & $ < $0.034& 11.4 & - \\
 HD\,205021 & Apr 24.59 &$ < $0.029 &9.8& 0.13 & Apr 22.60 &$ < $ 0.033 &11& 0.97 & Apr 22.56 & $ < $0.034& 11.4 & 0.96 \\
 HD\,163472 & May 05.27 &$ < $0.033 &11& 0.47 & Apr 28.29 &$ < $ 0.019 & 6.3& 0.56 & Jun 17.18 & $ < $0.033& 11.2 & 0.27 \\
 ALS\,9522 & May 23.32 &$ < $0.027 &9& - & Apr 26.35 & 0.080$ \pm $0.012 &6&  & May 13.31 & $ < $0.030& 10.0& - \\
 HD\,156424 & -&- &- & - & May 12.31 & 0.381$ \pm $0.015 &6& 0.06 & Jun 26.21 & 0.494$ \pm $0.028& 14.0& 0.70 \\
 HD\,3360 & Mar 08.93& $ < $0.029& 9.6& 0.02 & Mar 28.01 & $ < $ 0.018& 6& 0.57 & Jul 11.63 & $ < $0.038& 12.6 & 0.24 \\
 HD\,58260 & Apr 25.92 & $ < $0.036 & 12& - & Apr 22.98 & $ < $ 0.042 & 14 & - & Jul 20.71& $ < $0.075& 25 & - \\

  \hline 
   \multicolumn{12}{l}{Notes: The upper limits on the flux densities are 3-$\sigma$ upper limits. All the observations were taken in the year 2014.}
\end{tabular}

\end{sidewaystable*}

\section{Results}
\label{results}

This survey resulted in a total of four detections in the X band. The detected stars are HD\,37742 (O9.5 Ib), HD\,47129 (O7.5 III), HD\,156424
(B2 V) and ALS\,9522 (B1.5 V). Three of the stars detected in the X band are also detected in the Ka band (i.e. all except for ALS\,9522).
Only one of them, HD\,37742, was also detected in the S band. 
The non-detections in the S band are probably due to free-free absorption. This is because, due to the wavelength-squared dependence of the free-free opacity \citep[e.g.][]{weiler}, the size of the radio photosphere increases with wavelength resulting in high free-free absorption at S-band frequencies compared to the X and Ka bands.
 

\subsection{Mass-loss rates}
\label{massloss}

In this section, we estimate the  mass-loss rates from our radio observations, and compare them with the mass-loss rates calculated from theoretical models which are based on free-free emission. This may allow us to understand the emission processes and  estimate the thermal versus non-thermal contributions to the radio emission. If the radio radiation is purely thermal, we would expect that the mass-loss rates estimated from observations to match those
 expected from theoretical models. If the radio emission has an important non-thermal contribution, then our estimates based on the thermal emission assumption are likely to be overestimates.

The theoretical mass-loss rates were taken from \citet{naze14}. They used the models of \citet{vink00} as a function of effective temperature, mass, luminosity and wind terminal velocity. The model assumes that the star is isolated, neglecting any effects of binarity. They assume that the atmosphere consists of H, He, C, N, O and Si. The model takes multiple scattering into account. 
The theoretical mass-loss rates correspond to those that the stars would have in the absence of a surface magnetic field. It treats the gas in non-LTE (Local Thermodynamical Equilibrium) with the transfer of radiation computed in an expanding atmosphere. The temperature structure is from the assumption of radiative equilibrium in a spherically symmetric LTE atmosphere. The  surrounding medium is also considered to be smooth. Thus, the realistic situation may be far from this ideal,  resulting in some uncertainties in theoretical mass-loss rates. However, these uncertainties are likely to change the mass-loss rate by about a factor of 3 \citep{krticka14,smith14}, much less than the discrepancies we report in table \ref{table4}.
 
We calculated the expected mass-loss rates from our X-band radio data,  assuming the observed radio emission is purely thermal. The free-free radio flux density $S_{\nu}$ in Jy is related to the stellar mass-loss rate $\dot{M}$ by:  \citep{bieging89}. 
\begin{equation}
\dot{M} = \dfrac{3.01\times 10^{-6} \mu}{Z (\gamma g_{\rm ff} \nu)^{1/2}} V_{\infty} S_{\nu}^{3/4} D^{3/2} M_{\odot} \rm yr^{-1}
\end{equation}
where $ \mu $ is the mean ionic weight in amu, $v_{\infty}$ is the terminal velocity in kms\,$ ^{-1} $, $D$ is the distance to the star in kpc, $\nu$ is the frequency in GHz, $\gamma$ is the mean number of electrons per ion, $Z$ is the rms charge per ion and $g_{ff} $ is the free-free Gaunt factor, given by:

\begin{equation}
g_{\rm ff} = -1.66+ 1.27 \log(T_{\rm wind} ^{3/2}/(Z\nu) )
\end{equation}

\noindent where T$ _{\rm wind} $ is the local temperature (in K) at the radio photosphere.

\begin{table}
\begin{footnotesize}

\caption{Theoretical and empirical mass-loss rates of the stars.}
\label{table4}
\begin{tabular}{ p{2.3cm} p{2.3cm} p{2.3cm}  }
\\
\hline
 Name & $\dot{M_{th}}$ $(\times 10^{-6} M_{\odot}$\,yr$^{-1}$) & $\dot{M_{obs}}$ $(\times 10^{-6} M _{\odot}$\,yr$^{-1}$) \\ 
\hline
 HD\,191612 & 0.79 & $<$ 0.85  \\
 NGC\,1624-2 & 0.16 & $<$ 3.71  \\
 HD\,47129 & 0.06 &  4.89 \\
 HD\,108 & 2.51 & $<$ 0.93  \\
 HD\,37742 & 1.12  & 1.25  \\
 HD\,57682 & 0.08 & $<$ 0.56  \\
 HD\,149438 & 0.02 & $<$ 0.02  \\
 HD\,66665 & 0.006 & $<$ 0.47  \\
 HD\,46328 & 0.03 & $<$ 0.06  \\
 HD\,47777 & 0.002 & $<$ 0.27  \\
 HD\,205021 & 0.002 & $<$ 0.03  \\
 HD\,163472 & 0.0003 & $<$ 0.04  \\
 ALS\,9522 & 0.01 & 0.67 \\
 HD\,156424 & 0.003 & 1.34 \\  
 HD\,3360 & 0.004 & $<$ 0.008  \\

\hline
 \multicolumn{3}{l}{Notes: $ \dot{M_{th}} $ is the theoretical mass-loss rate  \citep{vink00} }

\end{tabular}

\end{footnotesize}

\end{table}

Table \ref{table4} shows the mass-loss rates derived using the above procedure. We also list their respective theoretical mass loss rates. We have listed only 15 of 18 stars, as only these stars have theoretical mass loss rates reported in the literature \citep{naze14}. In the case of the stars that are in binary systems, the wind from both the stellar components have not been considered for evaluating theoretical mass-loss rates, which could create some discrepancy 
in the correct estimation of the mass loss rate. In 3 of the detected stars, we find that for HD4\,7129, HD\,156424, and ALS\,9522, the mass-loss rates derived from radio observations are many times higher than those expected from the theoretical models. This could suggest  a significant contribution of radio emission by various mechanisms, e.g. non-thermal radio emission or radio emission from colliding wind binary. For HD\,37742, the mass-loss rate obtained from radio observations matches the value obtained using  the
theoretical model. This result could be evidence for a thermal origin of its  radio emission.  For the non-detected stars, upper limits on estimated mass-loss rates (from 3$\sigma $ upper limits on the radio flux)  are consistent with the theoretical mass-loss rates for those stars, i.e. the theoretical model predicts radio fluxes that are below our detection threshold.


\subsection{Detections}
\label{result}

Here we discuss the properties of the detected stars and the nature of their radio emission.

\subsubsection{HD\,47129}

HD\,47129  \citep[Plaskett's star;][]{plas22} is a massive binary system composed of O8 III/I and O7.5 III stars with an orbital period of 14 days \citep{linder08}.
 A strong, organized magnetic field was detected in the rapidly rotating secondary component  \citep[with a dipole polar strength of $\sim 2800$~G, ][]{grunhut13}, while the primary is not known to host a magnetic field. \citet{grunhut13} show that the magnetic star has evidence of magnetic field inversion. This stellar system has been shown
 to be chemically peculiar, and is a hard, luminous and variable 
source of X-rays  \citep{linder08}. The combination of strong magnetic field and rapid rotation leads to the existence of a centrifugal magnetosphere \citep[c.f.][]{petit13} surrounding the secondary star \citep{grunhut13}. This is in contrast to all other known magnetic O-type stars that host dynamical magnetospheres.
HD\,47129 is proposed to be a post-mass transfer (post-Roche lobe overflow) system, which could explain the observed mass-luminosity mismatch, chemical peculiarity and rapid rotation of the secondary star { \citep{bag92,linder08}}.   
 
 To phase the observations of Plaskett's star, we used the rotational ephemeris for the magnetic secondary star reported by Grunhut et al (in prep):

\begin{equation}\label{hd47129ephem}
HJD = 2455961.000+ 1.21551\cdot E.
\end{equation}

However, there may be  evidence for a period change  (Grunhut et al, in prep). The rotation period of this star was determined using a combination of magnetic, spectroscopic and photometric measurements. Grunhut et al. (in prep) obtained 63 magnetic observations of the system, which they combined with equivalent width measurements of the He II ($\lambda =4686 $A$^{0}$) and H$\beta$ lines and CoRoT photometry, to infer the rotation period of the magnetic component, P=1.21551d. This period is firmly established, and phase zero corresponds to the phase of maximum longitudinal magnetic field. Radio emission is clearly detected in the X and Ka bands at phases 0.47 and 0.13 respectively, according to Eq.~\ref{hd47129ephem}. 
We cannot unambiguously determine the spectral index of the source using the observations obtained in the various frequency bands as they correspond to different phases. However, due to the larger bandwidth of each band, we attempted to obtain the intraband spectral index. For this purpose, we calculated continuum images using the first half and second half of the band, and flux densities were determined. 
The spectral index is calculated to be $  -1.2\pm 1.0$ between sub-bands of X band, i.e. between central frequencies of 9\,GHz and 11\,GHz. 
The value of the spectral index in the sub-bands of the Ka band, i.e. between central frequencies of 31 and 35\,GHz, is $   0.0 \pm $2.5.
The observed mass-loss rate in this case is roughly two orders of magnitude larger than the theoretical mass-loss rate. This indicates that radio emission may come from  different mechanisms. This star is in a close binary system (orbital period $\sim$ 14 days). This radio emission can be due to thermal free-free emission from the ionized medium surrounding the individual components of the binary and non-thermal radio emission from centrifugal magnetosphere surrounding the magnetic secondary star . In addition, in this close binary system, interacting
winds of the binary system can also give rise to thermal as well as
non-thermal radio emission.

\subsubsection{HD3\,7742}

HD\,37742 ($\zeta$~Orionis A) is the closest O-star in our sample. 
It is  an O-type binary star is one of the belt stars in the constellation Orion.  It was shown to be a binary system   composed of a O9.5I supergiant and a B1IV star with an orbital period of 2687 days \citep{hum13}. \citet{bouret08} determined the primary component $\zeta$~Ori Aa to be a 40\,$M_{\odot} $ star with a radius of 25 $R_{\odot} $ at an age of $5-6$\,Myr, showing no surface nitrogen enhancement and losing mass at a rate of about $2\times 10^{-6}\,M_{\odot} {\rm yr}^{-1}$. The magnetic field of the primary star is the weakest magnetic field ever detected on a massive star ($\sim $140 G, \citealt{blazere15}). Magnetic field is not detected on its companion $ \zeta $ Ori Ab with an upper limit of $ \sim $300\,G. Because the magnetic field and the rotation rate are weak and the stellar wind is strong, $ \zeta $ Ori Aa does not host a centrifugal magnetosphere. However, it may host a weak dynamical magnetosphere .
Although this star is also well known for its prominent X-ray emission, \citet{cohen14} found that the emission resembles that of a non-magnetic star in the X-ray domain. This is consistent with a negligible magnetospheres surrounding $\zeta$~Ori Aa.  

To phase the observations, we have used the rotational ephemeris given by \citet{bouret08}: 
\begin{equation}
HJD = 2454380.0 + 6.829 \cdot E
\end{equation}

HD\,37742 is detected in the S, X and Ka bands. The star has higher flux density in the Ka band as compared to the S and X bands. The intra-band spectral index 
was found to be $ 0.3\pm 0.3$ in the S-band, $ 0.5\pm 0.2$ in the X-band and $ 0.7\pm 0.4$ in the Ka band. The positive spectral index in all bands suggest that the emission has a significant contribution from thermal emission. This star was previously observed by \citet{ll93} at 3.6\,cm and 6\,cm. The reported flux densities are 0.89 $\pm $ 0.04\,mJy at 3.6\,cm and 0.7\,mJy at 6\,cm. They concluded that the emission from this star is thermal using the slope for the spectrum and the absence of polarization. Comparison with the observed H$\alpha $ emission also suggested that the star does not exhibit significant non-thermal emission at cm wavelengths { \citep{ll93}}. 
In addition, this is the only detected star for which the values of the observational and theoretical mass-loss rates are consistent with each other, again suggesting a thermal origin of the radio emission. This may indicate that its magnetosphere - expected to be very small - is not contributing significantly to the radio emission from this star.  Though, this star is in a binary system, longer orbital period ($\sim $2687\,days) suggests that winds of the individual stars may not be interacting. So, the radio emission is not likely to come from wind colliding region.

\begin{figure}
  \centering
  \includegraphics[width=1.05\linewidth]{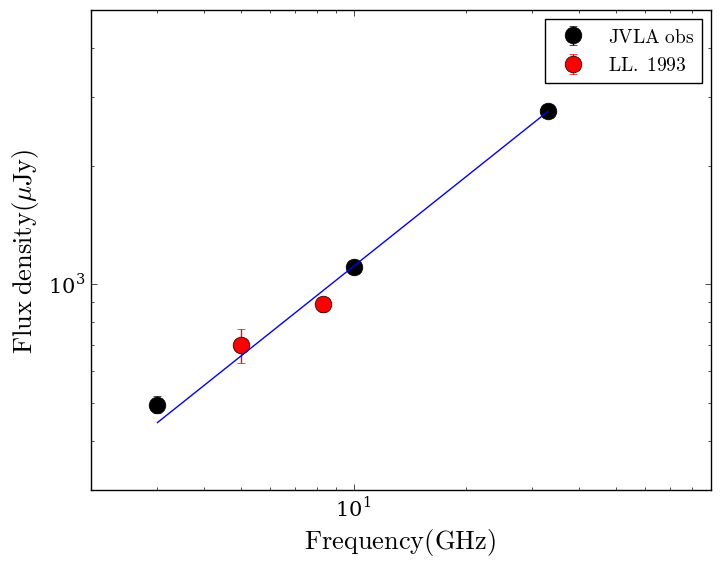}
  \caption{Log-log plot of flux density versus frequency for HD\,37742. Flux densities at 5\,GHz and 8\,GHz were taken from \citep{ll93}. Blue line is a power law fit to the observations with a spectral index  $\sim 0.7$, indicating thermal emission.  The JVLA data points have error bars but the error bars are smaller than the 
  symbol size. The 8 GHz data from \citet{ll93} does not have 
  error bars.}
  \label{hd37sp}
\end{figure}

Since thermal emission is likely to be independent of the rotational phase of the star, we can plot the flux densities at various frequencies irrespective of their phases. Fig \ref{hd37sp} shows the flux density versus frequency in the S, X and Ka bands. We have also included the flux densities at 5 GHz and 8.3 GHz from \citet{ll93}. The spectral index is found to be $ 0.76 \pm 0.03 $, suggesting that the emission is thermal. The spectral index can be used to infer information about the steepness of the density gradient. A spectral index of $\sim 0.6 $ is expected in the case of a stellar envelope in which the electron density ($n_{e}$) is proportional to $r^{-2}$, starting from the photospheric radius up to infinity \citep{olnon75, panagia75}. Assuming a smooth density profile, we can obtain the density structure around the star, which is found to be $n_{e} \propto r^{-2.2}$.  However, a clumpy medium can also change the
value of the spectral index.

\subsubsection{ALS\,9522}

ALS\,9522 (W601) is a B1.5V star and a member of the open cluster NGC\,6611 \citep{dufton06}.  This star has been classified as a pre-main sequence star and is an evolutionary progenitor of a main sequence B star \citep{mar07}.  A magnetic field was detected in this star by \citet{alecian08}, implying a dipole field of roughly 4\,kG \citep{petit13}. The existence of such a magnetic field in a pre-main sequence star supports the fossil field hypothesis, which supposes that the magnetic field is a relic from the field present in the interstellar medium from which the star was formed. The star is also known to be a X-ray emitter \citep{guar12}. This star was detected in the X-band, but was not detected in the S and Ka bands. No ephemeris for this star is available, hence we could not calculate the phases corresponding to the different observations. 

However, the longitudinal magnetic field observations of \citet{alecian08} show strong variability, suggesting that rotational modulation will be significant. 
 
Out of the detected stars at X-band, this is the only star which is not detected at the Ka band, which may be indicative of  non-thermal emission.
This is supported by the fact that the observational estimate of the mass-loss rate is roughly 2 orders of magnitude larger than the theoretical rate

\subsubsection{HD\,156424}

HD\,156424 is a He-strong B2V star belonging to the Sco OB4 association \citep{khar04}. A magnetic field was detected by \citet{alecian14} in the MiMeS survey, Further monitoring by \citet{shultz} identified the system as a radial velocity variable and candidate SB1 star, and allowed the derivation of a dipole magnetic field strength of 5.4\,kG. Of the 4 targets detected in the radio, this is the only star that has not been detected in X-rays \citep{naze14}. 
They derived an upper limit on the $0.5-10$\,keV X-ray flux of 
$\log(L_{\rm X})<29.8$\,erg\,s$^{-1}$.  \citet{asif14} carried out 2D MHD simulations to examine the effects of radiative cooling and inverse Compton
cooling on X-ray emission  in magnetic massive stars with radiatively driven stellar winds. 
In their semi-analytic scaling analysis, they estimated the X-ray flux for this star to be
$\log(L_{\rm X})=29.7$ erg $^{-1}$, which is compatible with the observational upper limits.

We have used the ephemeris of \citet{shultz} to phase our observations. 

\begin{equation}
HJD = 2456126.664 + 2.87233 \cdot E 
\end{equation}

The rotation period of this star was determined using a combination of magnetic, spectroscopic and photometric measurements. Shultz et al. (in prep) obtained 11 magnetic measurements of HD\,156424 which they combined with a dozen high resolution spectra. Both magnetic and Halpha EW measurements combine to suggest period of 2.8721\,d. The zero-point corresponds to the largest unsigned longitudinal field.
This star is detected in the X and Ka bands, but it  was not observed in S-band. The flux density at 33 GHz is higher than the flux density at 10\,GHz. However, this could reflect the slope of the spectral energy distribution, or it might be due to flux variability between different phases. In order to obtain an estimate of the spectral index, we have evaluated the flux densities using the first half and second half of the X and Ka bands.  Evaluated intra-band spectral indices are  $ -0.5\pm 0.5$  and $ -1.3\pm 2.0$ at the X and Ka bands respectively. The observational estimate of the mass-loss rate for this star is
nearly 500 times the theoretical rate, strongly implying an important non-thermal contribution to its emission.

\section{Discussion}
\label{discussion}

In this section, we examine relationships between the measured radio properties of the detected and non-detected magnetic stars and their magnetospheric properties. 


\begin{figure}
\centering
\includegraphics[width=1.0\linewidth]{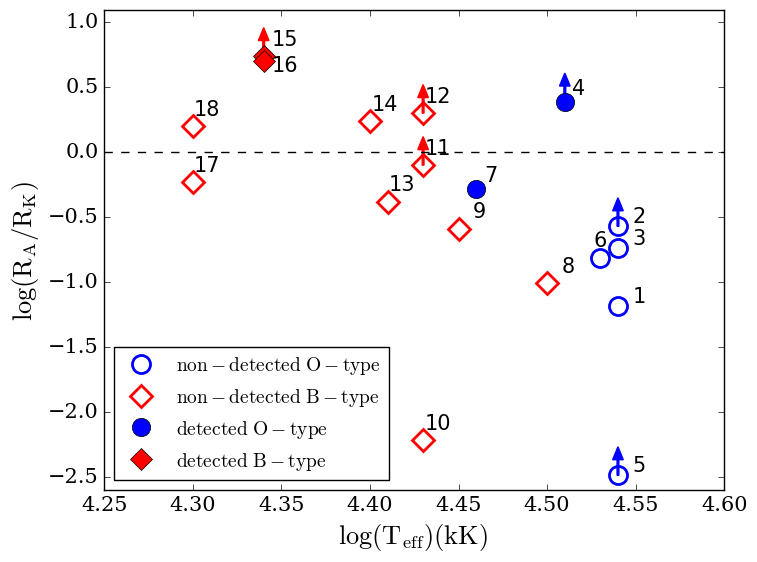}
\includegraphics[width=1.0\linewidth]{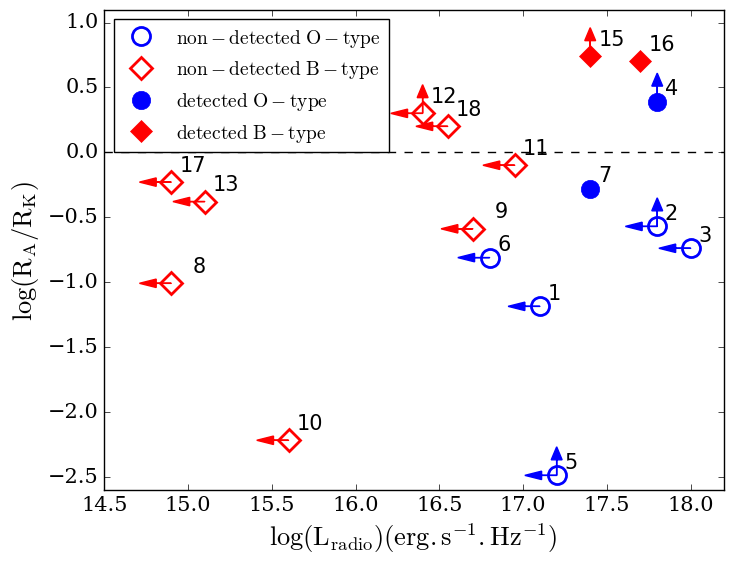}
\caption{Location of magnetic B-type and O-type stars in a log-log plot of $R _{\rm A} / R_{\rm K}$  vs. effective temperature (top panel) and radio luminosity (bottom panel). The horizontal line in both plots shows the division between centrifugal magnetospheres (area above the line) and dynamical magnetospheres (area
below the line). Blue circles indicate O-type stars and red diamonds indicate B-type stars. Filled symbols indicate radio-detected stars. The labels refer to ID sequence number listed in Column 1 of Table \ref{table1}. }
\label{fig:rarktl}
\end{figure}%

\begin{figure}
\centering
\includegraphics[width=1.0\linewidth]{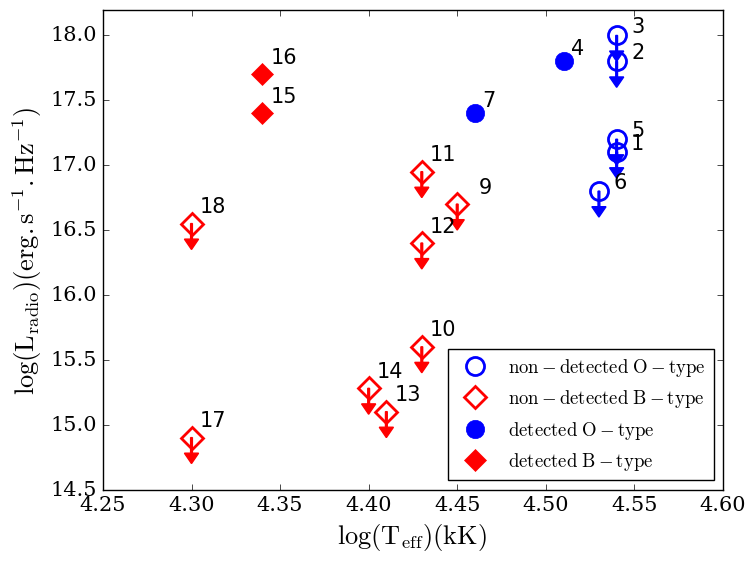}
\includegraphics[width=1.0\linewidth]{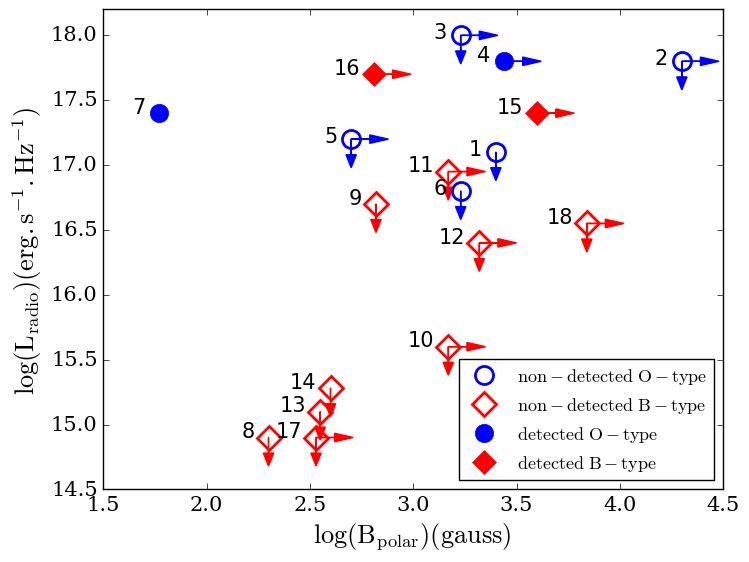}
\caption{Location of magnetic O-type and B-type stars in a log-log plot of radio luminosity vs. effective temperature (top) and dipole magnetic field strength (bottom). Blue circles indicate O-type stars and red diamonds indicate B-type stars. Filled symbols indicate radio detections. Upper limits on luminosities were calculated from $3\sigma$ upper limits on flux densities. The labels refer to ID sequence number listed in Column 1 of Table \ref{table1}.}
\label{fig:lumteff}
\end{figure}%


Fig \ref{fig:rarktl} shows the location of the observed stars in a log-log plot of $R _{\rm A}/R_{\rm K}$  vs the effective temperature (upper frame) and the radio luminosity at 10 GHz (lower frame). 
The horizontal line in both the plots at $R_{A}=R_{K}$ shows the theoretical division between stars hosting centrifugal magnetospheres (upper region) and dynamical magnetospheres (lower region). 
The distance of a target above the horizontal line characterizes the radial extent of the centrifugal support, and serves as a proxy for the volume of the centrifugal magnetosphere. We note that all of the detected stars host centrifugal magnetospheres except for HD\,37742, which likely hosts a dynamical magnetosphere \citep{blazere15}.  Comparison of mass-loss rates (\S \ref{massloss}) and spectral index  (\S \ref{result}) suggests that the emission from HD\,37742 is thermal. It appears that non-thermal radio emission seems to favour centrifugal magnetospheres. This is qualitatively expected, as electrons can be more easily accelerated to relativistic speeds in the high density plasma of a centrifugal magnetosphere to produce non-thermal emission \citep{trigilio04}.

Radio luminosities plotted in Fig \ref{fig:rarktl} are calculated from the radio flux densities in the X band. Upper limits on radio luminosities were calculated from $3\sigma$ upper limits on flux densities for the non-detections. We attempted to determine if the radio luminosity depends upon the stellar properties such as effective temperature or magnetic field. 
 Fig \ref{fig:lumteff}  shows the log-log plot of radio luminosity vs. the effective temperature (upper frame) and the dipole magnetic field (lower frame).   Radio luminosities show no correlation with effective temperature, nor do they show any relationship with magnetic field strength. Among the radio-detected stars, HD\,37742 has very weak magnetic field. The remaining detected stars have magnetic fields of similar strength to those of the  population of undetected stars. 

We can compare the radio detections of our sample of magnetic stars to the VLA survey of 88 non-magnetic OB stars carried out by \citet{bieging89}. Those authors detected a total of 14 sources at 6 cm. They observed non-thermal emission in 6 of the stars and free-free emission in 8 stars. The criterion used for establishing the nature of radio emission is the radio spectrum. They  looked at the general shape of spectrum at different epochs, and found that non-thermal emission was more efficient in optically luminous stars.  However, we did not find any correlation of detections with their optical luminosities. 


\citet{naze14} studied the X-ray emission of magnetic OB stars, which contains all the stars in our sample. They found that two O
stars (HD\,37742 and HD\,47129) show distinct X-ray properties, indicating that the main origin of the X-ray emission is most probably non-magnetic. Both of these O stars are detected in radio bands. 

We discuss the results and properties of O-type stars and B-type stars from our sample in the following subsections in detail.

\subsection{O-type stars}

Two out of seven O-type stars were detected in our VLA observations. All the O-type stars have dynamical magnetospheres except for radio detected Plaskett's star (HD\,47129). This is not surprising, as all the O-type stars have strong stellar winds due to their high luminosities, which implies a rapid stellar spin down due to magnetic braking. As Plaskett's star is in a close binary system, it was probably  spun up due to binary interaction \citep[e.g.][]{grunhut13}. 

The other radio detected star HD\,37742 may host a weak magnetic field,
and
may not have much contribution from the magnetosphere. This 
has been confirmed as the comparison of mass-loss rates (\S \ref{massloss}) and spectral index  (\S \ref{result}) strongly suggest that the emission from HD\,37742 is thermal.  
 
On the contrary, the mass loss rate of HD\,47129 is two orders of magnitude larger than the theoretical estimate. 
 This is a close binary system consisting of O-type stars, with the 
secondary having a well ordered large scale magnetic field. In addition to thermal free-free emission from the ionized medium
surrounding the individual components of the binary, 
a centrifugal magnetosphere surrounding the magnetic secondary star
could be a source of non-thermal radio emission.
In addition, in this close binary system, interacting winds of the binary can also give rise to thermal as well as non-thermal radio emission \citep{stevens, pittard10}. Thus the
 total radio emission in a close binary will be due to thermal components from individual stars, as well as thermal and non thermal component from the colliding wind region \citep{pittard10}. 
\citet{stevens} investigated the total thermal flux from colliding wind binaries and found that it could be about 50\% higher than that expected from the free-streaming stellar wind alone.
MHD simulations by  \citet{pittard10} have shown that the thermal component due to the colliding wind will start to dominate at high frequencies $\sim 50$~GHz, where as the non-thermal component coming from the magnetosphere is likely to be absorbed. At lower frequencies, emission from individual stars will be important. 

However, if the  stars in the binary system have large separation, the wind interaction is 
not likely to be effective. For example, HD\,37742  is in a binary system, but with a much longer orbital period ($\sim $2687 days) and
 radio emission is explained well by the free-free emission of the
 ionized wind of the star, without any need to incorporate wind interaction.  
 
Binarity remains a fundamental question regarding the non-thermal radio emission from O-type stars. \citet{vanloo05} concluded that non-thermal radio emission from O stars cannot be explained by wind-embedded shocks, and is likely to be caused by a colliding wind shock. That would imply that all O stars with non-thermal radio emission should be members of binary or multiple system. Of the 16 currently-known non-thermal radio emitting stars, 15 are confirmed binaries \citep{vanloo06}, however none of them is known to be a magnetic star.  In our sample, two of the detected O-type stars are known to be binaries.  However, only HD\,47129 could be in the limit of being classified as colliding wind binary, with an orbital period of $\sim$14 days. Other binary stars have much longer orbital periods, indicating large binary separations.  Simultaneous observations of HD\,47129 at different frequencies are required to understand if the emission is thermal or non-thermal, and to investigate any relationship with the binary properties of the system. Observations on a larger sample of O-type stars may shed more light on correlation of binarity with non-thermal radio emission.

The spectral index measurements  may also be able to throw light on radio emission mechanism.  
We estimate a spectral index $\alpha$ ranging from $-0.2$ to $-2.2$, between 9--11 GHz frequencies. 
If the radio emission from HD\,47129 is purely non-thermal,
the lowest expected flux density at 3\,GHz is about 0.26\,mJy using $\alpha=-0.2$.  \citet{linsky92}   estimated the 
minimum flux density to be about 50\% of the maximum (typically for  MCP stars)  for  oblique rotators, if the radio emission is modulated.
Thus one should have expected a flux density of $\sim 0.13$\,mJy at 10\,cm, easily detectable at JVLA sensitivity.
Thus the non-detection of HD\,47129 at 10cm suggests that a significant contribution may come from the 
wind-wind interaction in this close binary system.


We have searched the literature for single O-type stars that are thermal radio sources from the literature. While their numbers are handful, their theoretical mass-loss rates are in good agreement with mass-loss rates derived from their radio fluxes \citep{bieging89,benaglia01,ll93}.  
 The radio luminosities of the detected O-type stars are similar to the upper limits of the non-detected O-type stars. Hence it is not clear whether the detected O-type stars exhibit radio luminosities that are quite different from those of the overall population of magnetic O stars (i.e. the high-luminosity tail of a broad distribution), or if they represent only marginally brighter examples of a more uniform distribution. 
  One surprising result is the lack of any detection of radio emission from NGC\,1624-2, the star with the strongest magnetic field of the sample (by nearly an order of magnitude). The observations of this star yield a $3\sigma$ upper limit that is similar to that of the other non-detected O-type stars. Strong absorption by wind created from mass-loss could explain this.

\subsection{B-type stars} 

Both the detected B-type stars, HD\,156424 and ALS\,9522, host a CM and lie above all the non-detected B-type stars in Fig \ref{fig:rarktl}. This implies that both the detected stars have a larger radial extent of centrifugal support of magnetospheric plasma compared to the non-detected stars.  

The detected B-type stars are significantly more radio luminous than the majority of the upper limits of the non-detected B-type stars.  In fact, for all but 3 of the undetected B-type stars, the radio luminosity upper limits are at least an order of magnitude below the detected stars. This implies that the radio emission properties of the detected stars are, statistically speaking, very different from those of the undetected stars.  

In the case of the detected B-stars, the implied mass loss rates are much higher than the theoretical estimates. Since the detected B stars are not known to be
in binaries, the excess flux is likely to be a contribution from the magnetosphere from the magnetic star.

 We can compare the radio detections of B-type stars in our sample to the VLA survey of magnetic chemically peculiar (MCP) stars carried out by \citep{linsky92}. They have detected a total of 16 out of 61 sources at 6\,cm. They found a wide range of 6\,cm radio luminosities for the detected Bp stars, with log(L$_{6 cm} $) $=$ 15.7-17.9. The detected B-type stars from our sample have radio luminosities 17.4 and 17.7 at 3\,cm. These values lie within the broad range of MCP stars. They have also found that the radio luminosities of MCP stars correlate positively with effective temperature and magnetic field strength. However, no such correlations were found for the B-type stars in our sample.


\section{Summary and Conclusions}
\label{summary}
We have initiated a systematic survey of the radio emission properties of the known magnetic OB stars, which includes observations at both high frequencies using the JVLA and at low frequencies using the GMRT. The project is ongoing. In this paper we have presented the results of JVLA observations of 18 magnetic O- and B-type stars with masses greater than $8~M_{\odot}$. The JVLA observations were taken at random rotational phases in the S, X and Ka bands.  
We have detected X band radio emission from 2 out of 7 magnetic O-type stars and 2 out of 11 magnetic B-type stars in our sample. The detected O-type stars, HD\,37742 and HD\,47129, are in binary systems.
 
The detected B-type stars, HD\,156424 and ALS\,9522, are not known to be in binaries. Two other O-type stars, which are known to be in (much longer-period) binary systems (HD\,108, HD,191612) were not detected.  Only HD,37742 is detected in the S band.   The general lack of detections of our targets in the  S-band is probably due to free-free absorption by the free-streaming stellar wind.
Three stars, HD,37742, HD\,47129 and HD\,156424, were detected in the Ka band. The radio flux of HD\,37742 and HD\,47129 in the Ka band is consistently higher compared with that in the S and X bands. This can be explained by a dominant contribution of thermal flux.

Mass-loss rates were estimated for the detected stars using X band radio flux densities and were compared with the expected mass-loss rates from the theoretical models. However, there are some caveats in theoretical prediction of mass-loss rates as
binarity and the presence of magnetic field were not taken into account. In addition, the theoretical estimates assume smooth density profiles, while clumping factors of order 10 are measured in O-type star winds. However, the
clumpiness of the medium is not expected to change the mass-loss rate
by more than a factor of 3 \citep{smith14},  much less than the discrepancy we note.
For HD\,37742, the theoretical estimate matches with that of observational one, suggesting that the radio emission is mostly thermal. The thermal nature of the HD\,37742 radio emission is also supported by the spectral index we measure. For the remaining 3 stars, mass-loss rates estimated from radio observations were orders of magnitude  higher compared to those predicted by theory. This may indicate significant contribution of the
 radio emission from other mechanisms than only
 thermal free-free emission. In magnetic stars, middle magnetosphere
 can give rise to non-thermal gyrosynchrotron emission. However, in case of binary systems, the stellar wind from both the stars may interact and 
 produce thermal and non-thermal emission. The detected B- stars in our sample are not known to be in binary system, thus the additional mass-loss rate is likely to have contribution from the non-thermal emission of the 
 magnetosphere. However,  HD\,47129 is a close binary system and emission from colliding stellar winds could play a significant role.

All the detected stars host centrifugal magnetospheres except for HD\,37742, which hosts a dynamical magnetosphere. However, the  radio emission of HD\,37742 is unambiguously thermal. This suggests that non-thermal radio emission seems to favour centrifugal magnetospheres. In addition, binary wind interactions may also play a role.

We were unable to evaluate the nature of emission or variability of radio flux as we have a adopted snapshot approach to identify the stars that are emitting radio radiation.  In order to understand the emission mechanisms and flux variability over the rotation period, we need to observe the detected stars simultaneously over all the frequencies. 

The detectability of the O-type stars seem to be sensitivity limited. With the upcoming SKA, we expect to achieve a significant sensitivity 
and detect a larger fraction of the O star sample. 


\section*{Acknowledgements}

 We thank referee for  very constructive comments, which helped improve
the manuscript tremendously.
P.C. acknowledges support from the Department of Science and Technology via SwaranaJayanti Fellowship award (file no.DST/SJF/PSA-01/2014-15).
G.A.W. acknowledges Discovery Grant support from the Natural Science and Engineering Research Council (NSERC) of Canada.
AuD acknowledges support by NASA through Chandra Award numbers GO5-16005X, AR6-17002C, G06-17007B and proposal 18200020 issued by the Chandra X-ray Observatory Center which is operated by the Smithsonian Astrophysical Observatory for and behalf of NASA under contract NAS8- 03060.
A.D.U. gratefully acknowledges support from the \textit{Fonds
qu\'{e}b\'{e}cois de la recherche sur la nature et les technologies}
, and DHC for NASA Chandra grants TM4-15001B and GO6-17007D. RHDT acknowledges support from NSF SI2 grant ACI-1339600 and NASA TCAN grant NNX14AB55G. 
The National Radio Astronomy Observatory is a facility of the National Science Foundation operated under cooperative agreement by Associated Universities, Inc. 












\bsp	
\label{lastpage}
\end{document}